\documentclass[fleqn,usenatbib]{mnras}

\usepackage{newtxtext,newtxmath}

\usepackage[T1]{fontenc}
\usepackage{tikz}

\DeclareRobustCommand{\VAN}[3]{#2}
\let\VANthebibliography\thebibliography
\def\thebibliography{\DeclareRobustCommand{\VAN}[3]{##3}\VANthebibliography}

\usepackage{graphicx}	
\usepackage{amsmath}	
\usepackage{xcolor}
\usepackage{caption}

\title[Carbon dredge-up and the Gaia WD CMD bifurcation]{Carbon dredge-up required to explain the Gaia white dwarf colour--magnitude bifurcation}

\author[Blouin, Bédard \& Tremblay]{
Simon Blouin$^{1}$\thanks{E-mail: sblouin@uvic.ca},
Antoine Bédard$^{2}$,
Pier-Emmanuel Tremblay$^{2}$
\\
$^{1}$Department of Physics and Astronomy, University of Victoria, Victoria, BC V8W 2Y2, Canada\\
$^{2}$Department of Physics, University of Warwick, Coventry, CV4 7AL, UK\\
}

\date{Accepted 2023 May 23. Received 2023 May 22; in original form 2023 March 15}

\pubyear{2023}

\begin{document}
\label{firstpage}
\pagerange{\pageref{firstpage}--\pageref{lastpage}}
\maketitle
\begin{abstract}
The {\it Gaia} colour--magnitude diagram reveals a striking separation between hydrogen-atmosphere white dwarfs and their helium-atmosphere counterparts throughout a significant portion of the white dwarf cooling track. However, pure-helium atmospheres have {\it Gaia} magnitudes that are too close to the pure-hydrogen case to explain this  bifurcation. To reproduce the observed split in the cooling sequence, it has been shown that trace amounts of hydrogen and/or metals must be present in the helium-dominated atmospheres of hydrogen-deficient white dwarfs. Yet, a complete explanation of the {\it Gaia} bifurcation that takes into account known constraints on the spectral evolution of white dwarfs has thus far not been proposed. In this work, we attempt to provide such a holistic explanation by performing population synthesis simulations coupled with state-of-the-art model atmospheres and evolutionary calculations that account for element transport in the envelopes of white dwarfs. By relying on empirically grounded assumptions, these simulations successfully reproduce the bifurcation. We show that the convective dredge-up of optically undetectable traces of carbon from the deep interior is crucial to account for the observations. Neither the convective dilution/mixing of residual hydrogen nor the accretion of hydrogen or metals can be the dominant drivers of the bifurcation. Finally, we emphasize the importance of improving theoretical models for the average ionization level of carbon in warm dense helium, which governs the shape of the diffusive tail of carbon and in turn the predicted amount of dredged-up carbon.
\end{abstract}

\begin{keywords}
equation of state -- stars: abundances -- stars: atmospheres -- stars: Hertzsprung--Russell diagrams  -- white dwarfs
\end{keywords}

\maketitle

\section{Introduction}
\label{sec:intro}
Since its second data release in 2018 \citep{gaiadr2_2018}, the {\it Gaia} mission has revolutionized our understanding of white dwarfs. It has enabled a one-order-of-magnitude increase of the number of known white dwarfs \citep{jimenez2018,gentile2019,gentile2021}, the construction of large volume-complete samples \citep{tremblay2020,mccleery2020,obrien2023}, the detection of the heat release signature of core crystallization \citep{tremblay2019}, the discovery of hypervelocity runaway white dwarfs \citep{shen2018}, the detailed characterisation of the white dwarf binary and merger populations \citep{kupfer2018,cheng2020,pala2020,torres2022}, and the identification of polluted white dwarfs with peculiar compositions \citep{hollands2021,kaiser2021,klein2021}.

\begin{figure*}
    \includegraphics[width=2\columnwidth]{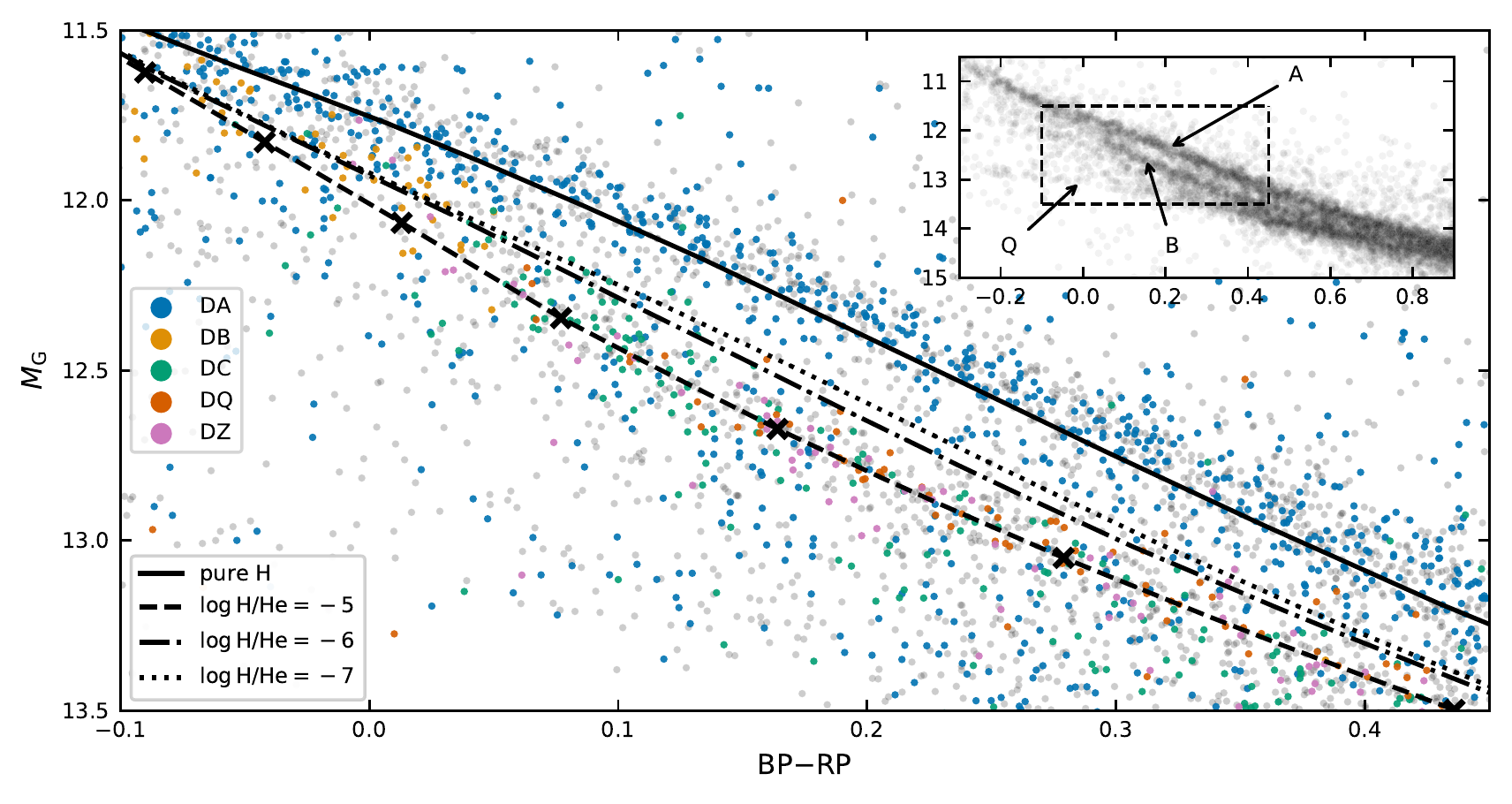}
    \caption{{\it Main panel}: The bifurcation in the {\it Gaia} white dwarf colour--magnitude diagram. Every circle represents a white dwarf in the Montreal White Dwarf Database (MWDD) 100~pc sample. The construction of this sample is based on {\it Gaia} magnitudes and parallaxes (see \protect\citealt{caron2023} for details). The colour of the circles represents the spectral class; non-spectroscopically confirmed white dwarfs are shown in grey. Note that only the ``primary'' spectral class is considered here (e.g., a DAZ white dwarf is displayed as a DA). Four $0.6\,M_{\odot}$ cooling sequences are shown: one for pure-hydrogen atmospheres (solid line) and three for helium-dominated atmospheres with traces of hydrogen (dashed, dash-dotted, and dotted lines). The upper feature traversed by the pure-hydrogen cooling track and almost exclusively populated by DA white dwarfs is known as the A branch; the lower feature traversed by the $\log\,{\rm H/He}=-5$ cooling track is the B branch. Note the paucity of objects in the gap between the two branches. For reference, crosses along the $\log\,{\rm H/He}=-5$ sequence mark effective temperatures in steps of 1000\,K, starting at 13{,}000\,K on the left and ending at 7000\,K in the lower-right corner. {\it Inset}: Zoom out on the 100~pc {\it Gaia} white dwarf colour--magnitude diagram. The A, B and Q branches are marked, and the region shown in the main panel is delimited by the dashed rectangle.}
    \label{fig:GaiaHRD_intro}
\end{figure*}

The presence of three distinct features in the {\it Gaia} white dwarf colour--magnitude diagram was one of the very first puzzles posed by the {\it Gaia} data \citep{gaiahrd2018}, and it remains only partially solved to this day. The first of these three features, the A branch, is well understood and simply represents the cooling sequence of canonical hydrogen-atmosphere DA white dwarfs with a mean mass of $\simeq 0.6\,M_{\odot}$ (\citealt{gaiahrd2018,bergeron2019} and Figure~\ref{fig:GaiaHRD_intro}). But uncertainties surrounding the exact nature of the two other features, the B and Q branches, persist. The Q branch has been interpreted as a pile-up structure resulting from the release of latent heat and gravitational energy from carbon--oxygen phase separation during core crystallization \citep{tremblay2019,bergeron2019}. However, the complete story appears to be more complicated as it has since become apparent that further energy sources, likely connected to additional phase separation processes, are needed to fully account for the observed overdensity \citep{cheng2019,bauer2020,blouin2020,blouin2021}. 

The B branch, clearly visible in Figure~\ref{fig:GaiaHRD_intro} under the A branch, is mostly constituted of cool ($T_{\rm eff} \lesssim 10{,}000\,$K) non-DA white dwarfs with helium-dominated atmospheres. These objects come in different flavours, including featureless DC white dwarfs (helium lines are not visible at these temperatures), DZ white dwarfs with spectral lines from rock-forming elements originating from accreted planetary material \citep{zuckerman2007}, and DQ white dwarfs displaying molecular carbon features due to the convective dredge-up of carbon from the deep interior \citep{pelletier1986,camisassa2017,bedard2022b}. Note that the presence of DA white dwarfs on and below the B branch is simply the signature of the high-mass tail of the DA mass distribution. In other words, the B branch is constituted of non-DA white dwarfs, but is ``contaminated'' by a number of high-mass DAs that lie below the A branch due to their smaller radii.

Atmosphere models of $0.6\,M_{\odot}$ white dwarfs with \textit{pure} helium atmospheres fail to replicate the B branch and instead go straight through the gap between the A and B branches \citep{gaiahrd2018}. However, if a significantly large trace of hydrogen is added to the helium-atmosphere models ($\log\,{\rm H/He} \equiv \log_{10}N({\rm H})/N({\rm He}) \simeq -5$), then the B branch can be successfully matched (\citealt{bergeron2019,kilic2020,gentile2021} and Figure~\ref{fig:GaiaHRD_intro}). Such small traces of hydrogen are too small to cause detectable Balmer lines, but they nevertheless provide additional electrons to the atmospheric gas that increase the He$^-$ free--free opacity (the main source of opacity in those stars, see Figure~18 of \citealt{saumon2022}) and in turn affect the predicted magnitudes. As noted by \cite{bergeron2019}, other electron donors (carbon or other metals) can in principle have a similar effect on the {\it Gaia} magnitudes, although to our knowledge this has not yet been explicitly explored in the context of the colour--magnitude diagram bifurcation.

The presence of hydrogen traces in helium-atmosphere white dwarfs on the B branch is a well-justified assumption given our current understanding of white dwarf chemical evolution. The small amount of hydrogen present in hydrogen-deficient white dwarfs should a priori stay at the surface of the star, floating above the thicker helium envelope. But as the white dwarf cools down, convection develops and the hydrogen layer can be mixed with the helium reservoir underneath, leading to helium-dominated atmospheres with traces of hydrogen. This can be achieved either through convective dilution, where convection in the helium envelope erodes the hydrogen layer at its base, or through plain convective mixing if a convection zone develops in the hydrogen layer and extends down to the helium mantle \citep{fontaine1987}. The exact thickness of the hydrogen layer determines which channel a particular hydrogen-deficient star will follow. Those with thick enough hydrogen layers will never mix; those with very thin layers will be mixed through convective dilution; and those with layers of intermediate thickness will undergo convective mixing \citep{rolland2018}.

While sufficiently large amounts of hydrogen can reproduce the bifurcation ($\log\,{\rm H/He}\simeq -5$) and while such levels can be achieved through known evolutionary processes, this scenario alone cannot fully explain the observed bifurcation. From the hydrogen-to-helium abundance ratio measurements or upper limits of DB(A) white dwarfs before the bifurcation \citep{koester2015,rolland2018,genest2019}, it is expected that a large number of helium-atmosphere white dwarfs on the B branch have $\log\,{\rm H/He} < -5$, with many having $\log\,{\rm H/He} < -6$ \citep[][Figure~8]{bedard2023}. As shown by the dotted line in Figure~\ref{fig:GaiaHRD_intro} and as noted by \cite{bergeron2019}, such a small trace of hydrogen is clearly insufficient to shift the cooling track out of the gap. So why are there no stars in the gap? Where are the descendants of DB(A) white dwarfs with $\log\,{\rm H/He} \lesssim -6$?

In this work, we perform population synthesis simulations coupled with state-of-the-art model atmospheres and time-dependent chemical evolution calculations to investigate this problem in details. We show that the predicted dredge-up of optically undetectable traces of carbon can naturally solve this conundrum, a solution initially proposed by \cite{bergeron2019} but that had not yet been modelled. We describe our numerical models and population synthesis simulations in Section~\ref{sec:models} and the main results of these calculations in Section~\ref{sec:results}. The discussion in Section~\ref{sec:discussion} explores two alternative explanations for the absence of stars in the gap and reviews important sources of uncertainties concerning the constitutive physics of our models. Finally, we conclude in Section~\ref{sec:conclusion}.

\section{Models and simulation assumptions}
\label{sec:models}
To investigate the cause of the {\it Gaia} bifurcation and establish the nature of B-branch white dwarfs, we perform population synthesis simulations that rely on model atmospheres and evolutionary calculations that follow the transport of chemical species in the envelope. We detail these models below.

\subsection{Model atmospheres}
We use the model atmosphere code described in \cite{blouin2018,blouin2018b} and references therein \citep[including in particular][]{dufour2005}. This code incorporates the latest constitutive physics for cool helium-dominated atmospheres. We make use of the grids of pure-hydrogen, He+H, and He+C model atmospheres that have been presented elsewhere \citep{blouin2019c,coutu2019}. In addition, we have generated 2250 new models to explore He+H+C compositions. This new grid spans surface gravities of $\log g ({\rm cm}\,{\rm s}^{-2}) =7.5$ to 8.5 in steps of 0.5 dex\footnote{For pure-hydrogen models, we have verified that using a grid with steps of 0.25 dex in $\log g$ yields, for a given mass and $T_{\rm eff}$, {\it Gaia} magnitudes that differ by no more than 0.001 mag from the values obtained with the coarser 0.5 dex resolution.}, effective temperatures of $T_{\rm eff}=6000\,$K to 16{,}000\,K in steps of 500\,K below 10{,}000\,K and in steps of 1000\,K above, carbon abundances from $\log\,{\rm C/He} = -8.0$ to $-4.0$ in steps of 0.5 dex, and hydrogen abundances from $\log\,{\rm H/He}=-7$ to $-3$ in steps of 1 dex. The {\it Gaia} magnitudes are computed from the synthetic spectra using the appropriate bandpasses \citep{gaiaedr3} and zero points calculated as in Section 5.4.1 of the {\it Gaia} DR3 documentation release 1.2.\footnote{\url{https://gea.esac.esa.int/archive/documentation/GDR3/}} Interstellar reddening is neglected as only objects within 100\,pc sample of the Sun will be considered when comparing model predictions to observations.

\begin{figure*}
	\includegraphics[width=\textwidth]{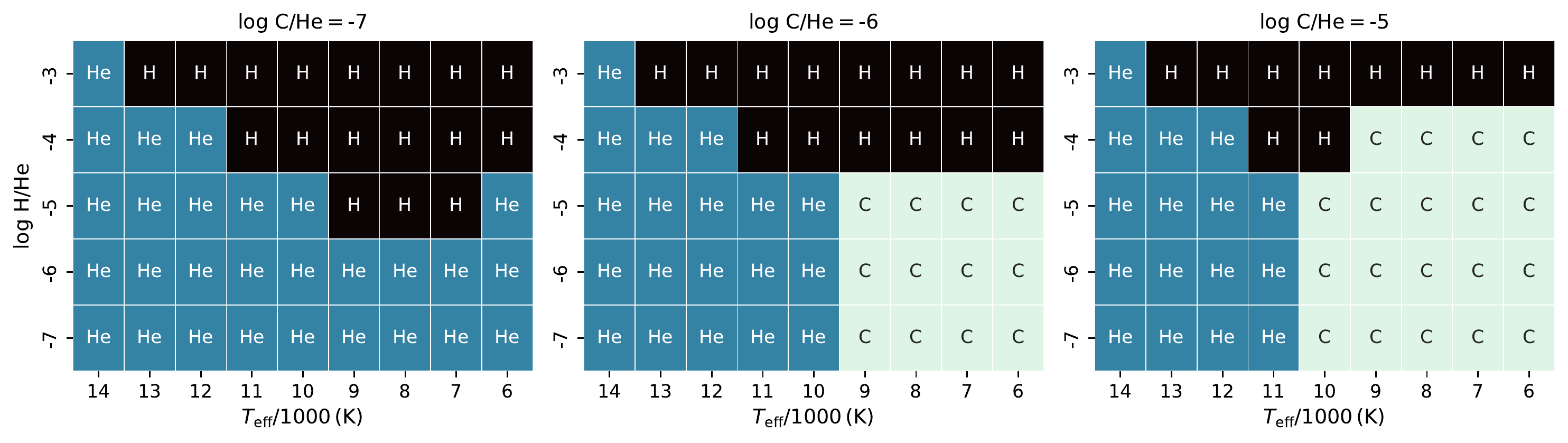}
    \caption{Main electron donor at the photosphere (Rosseland mean optical depth of 2/3) of  He+H+C atmosphere models as a function of the hydrogen content (vertical axis), effective temperature (horizontal axis), and carbon abundance (different panels). A surface gravity of $\log g=8$ is assumed.}
    \label{fig:electron_donor}
\end{figure*}

Figure~\ref{fig:electron_donor} shows the main electron donor (ionized hydrogen, helium, or carbon) at the photosphere of our helium-dominated model atmospheres as a function of the composition and effective temperature. Ionized helium always supplies the majority of the electrons at high temperatures and also dominates at low temperatures when the hydrogen and carbon concentrations are inferior to about one part per million. Provided that their concentrations are high enough, hydrogen and carbon dominate the electron budget at low temperatures, where helium remains essentially neutral. This reflects the fact that helium has a higher ionization potential (24.6\,eV) than hydrogen (13.6\,eV) or carbon (11.3\,eV). As explained in Section~\ref{sec:intro}, He$^-$ free--free absorption is the main opacity source in helium-atmosphere white dwarfs. As the intensity of He$^-$ free--free absorption directly depends on the free electron density, the heat maps of Figure~\ref{fig:electron_donor} effectively track the conditions under which hydrogen and carbon have a large impact on the spectral energy distribution. Green (black) cells correspond to conditions where carbon (hydrogen) will significantly affect the colours of the star. Of course, carbon and hydrogen can also have a non-negligible impact on the spectral energy distribution when helium is the main electron donor (blue cells).

\begin{figure}
    \includegraphics[width=\columnwidth]{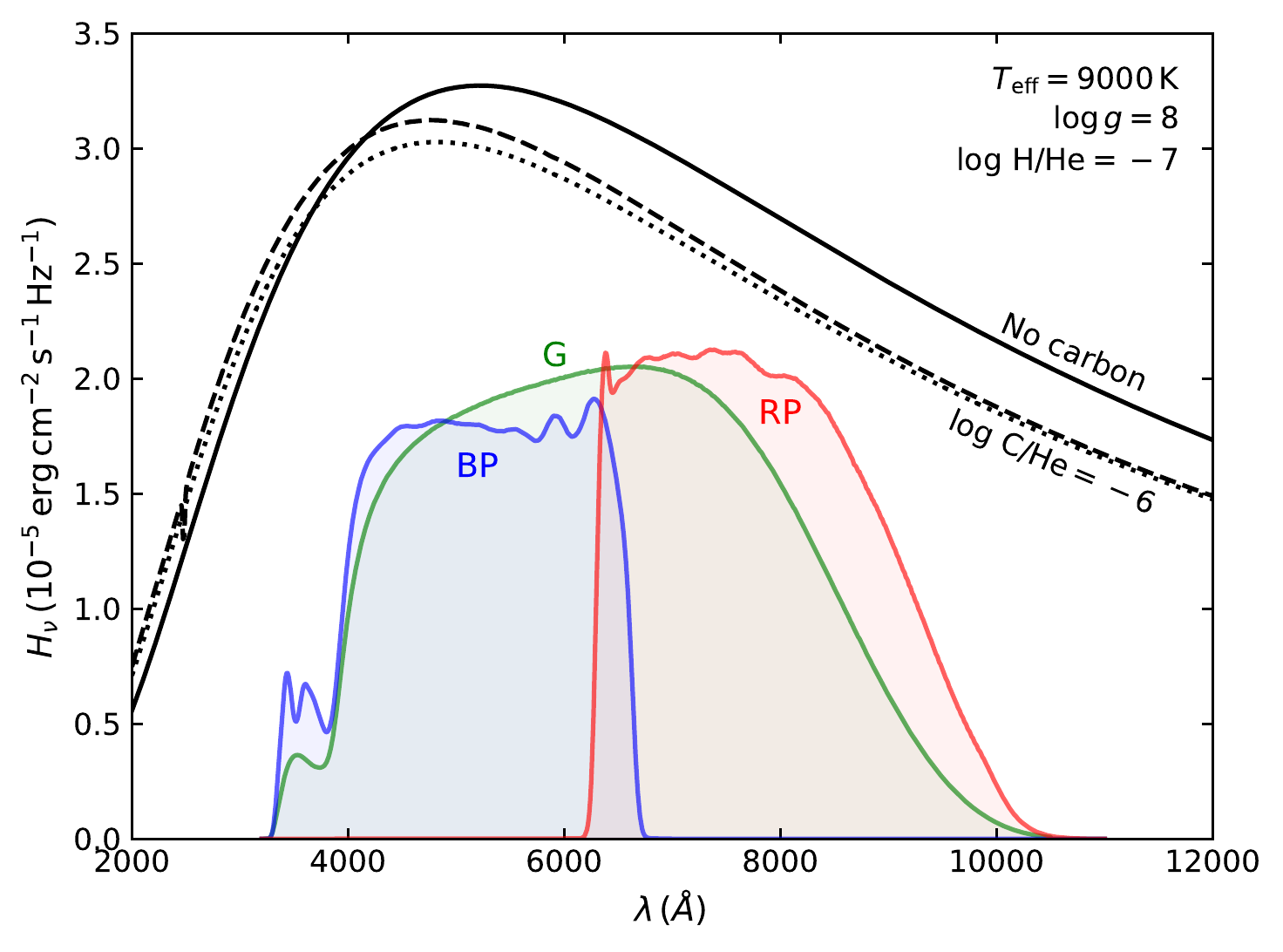}
    \caption{Synthetic spectra of a 9000\,K hydrogen-deficient white dwarf without any carbon (solid black line), with a trace of carbon (dashed black line), and with a trace of carbon but neglecting carbon opacity sources (dotted black line). The {\it Gaia} passbands are shown for reference (arbitrary scaling).}
    \label{fig:SED_demo}
\end{figure}

We compare in Figure~\ref{fig:SED_demo} the synthetic spectra of a $T_{\rm eff}=9000\,{\rm K}$ hydrogen-deficient white dwarf without carbon pollution and with a small trace of carbon ($\log\,{\rm C/He} = -6$, not spectroscopically detectable in the optical). The electrons provided by ionized carbon dominate the free electron budget (Figure~\ref{fig:electron_donor}), enhancing He$^-$ free--free absorption and making the star appear fainter in the {\it Gaia} $G$ band (here, $M_G$ increases by 0.08 mag) and bluer in BP--RP (BP--RP decreases by 0.08 mag). It is remarkable that such a small trace of carbon can change the spectral energy distribution so noticeably. Note that for these atmospheric parameters the impact of the carbon opacities is very small compared to the impact of the He$^-$ free--free absorption increase (i.e., there is only a small difference between the dotted and dashed lines in Figure~\ref{fig:SED_demo}).

\subsection{Evolutionary models}
\label{sec:stelum}
The atmosphere model calculations detailed in the previous section are oblivious to the chemical evolution of the star. They predict the emerging flux for a fixed atmospheric composition. While white dwarfs with thick hydrogen envelopes maintain their hydrogen-dominated atmosphere throughout their evolution, those with thinner hydrogen layers ($M_{\rm H} / M_{\star} \lesssim 10^{-6}$) can see significant changes to the chemical makeup of their atmospheres. This effect must be considered in our population synthesis simulations. As outlined in Section~\ref{sec:intro}, a first way by which a white dwarf with a thin hydrogen layer can change its photospheric composition is through the mixing of hydrogen with the helium envelope. This process has been investigated in great details both empirically \citep{sion1984,bergeron1997,bergeron2001,tremblay2008,koester2015,limoges2015,rolland2018,blouin2019c,ourique2019,cunningham2020,mccleery2020,lopez2022,caron2023} and through evolutionary calculations \citep{koester1976,dantona1979,macdonald1991,rolland2018,rolland2020,bedard2022a,bedard2023}. As detailed in Section~\ref{sec:assumptions}, we can build upon this body of work to include this aspect of chemical evolution into our population synthesis simulations and no new evolutionary calculations are required.

\begin{figure}
	\includegraphics[width=\columnwidth]{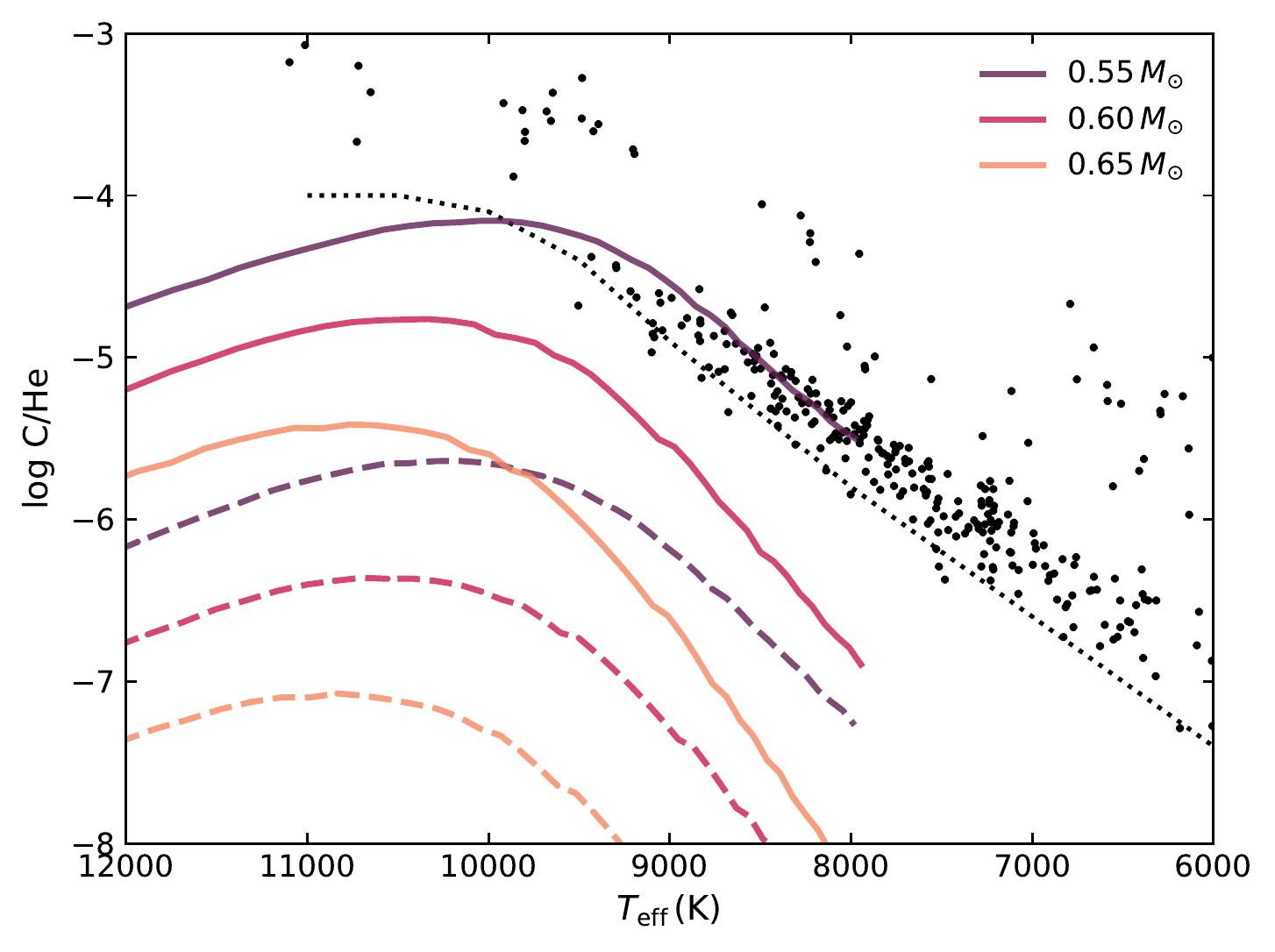}
    \caption{Photospheric carbon abundance as a function of effective temperature. The black circles correspond to carbon abundances measured in DQ white dwarfs by \protect\cite{coutu2019} and \protect\cite{blouin2019d}. The dotted line indicates the minimum abundance required for atomic or molecular carbon features to become detectable (assuming there are no other metals in the atmosphere, see \citealt{blouin2022}). The solid and dashed lines show the carbon abundances predicted by the evolutionary models of \protect\cite{bedard2022b} for different masses (see legend) and different initial carbon mass fractions in the envelope of the PG~1159 progenitor (solid lines correspond to $X_{\rm C}^{\rm init}=0.6$ and dashed lines to $X_{\rm C}^{\rm init}=0.2$).}
    \label{fig:DQseq}
\end{figure}

A second way through which chemical evolution operates in hydrogen-deficient white dwarfs is the convective dredge-up of carbon from the deep interior. Previous studies have mostly focused on explaining the observed $T_{\rm eff}-{\rm C/He}$ relation for DQ white dwarfs (Figure~\ref{fig:DQseq}). Here we are also interested in the dredge-up of small quantities of carbon that fail to produce detectable carbon lines or molecular bands but nevertheless affect the white dwarf's spectral energy distribution through an enhancement of He$^-$ free--free absorption. This scenario has not been extensively studied and as a result we need to perform new evolutionary calculations to properly include it in our population synthesis. The existence of such ``DQ-manqué'' stars can be inferred from the fact that the observed DQ $T_{\rm eff}-{\rm C/He}$ relation closely follows the carbon visibility limit (dotted line in Figure~\ref{fig:DQseq}). This suggests that DQ white dwarfs represent only the tip of the carbon abundance distribution among hydrogen-deficient white dwarfs, and that helium-dominated DC white dwarfs can in fact have significant (but optically spectroscopically invisible) traces of carbon in their atmospheres. Indeed, some DC white dwarfs have carbon traces detectable only in the ultraviolet \citep{weidemann1995}. Moreover, this interpretation is supported by evolutionary models, which predict that the magnitude of carbon pollution is highly sensitive to a number of stellar parameters, in particular the stellar mass and the carbon content of the PG~1159 progenitor. Consequently, the white dwarf mass distribution and the PG~1159 abundance distribution are expected to produce a broad range of carbon abundances in cool helium-dominated white dwarfs, with most of this range falling below the detection limit and thus giving rise to DC stars (see \citealt{bedard2022b} and the coloured lines in Figure~\ref{fig:DQseq}). As we have seen in Figure~\ref{fig:SED_demo}, such spectroscopically invisible traces of carbon can easily affect the star's spectral energy distribution.

To determine the carbon abundances to use in our population synthesis simulations, we performed element transport calculations using the STELUM evolutionary code \citep{bedard2022a}. These calculations are nearly identical to those described in \cite{bedard2022b}. Briefly, we evolve complete stellar models, solving the stellar structure and time-dependant species transport equations self-consistently. Gravitational settling, chemical diffusion, thermal diffusion, and convective mixing (including the effect of convective overshoot) are taken into account. Each sequence starts with a $T_{\rm eff}=90{,}000\,$K PG~1159-like progenitor made of an homogeneous C-O core surrounded by an homogeneous He-C-O envelope of mass $10^{-2}M_{\star}$. Calculations were carried out for three different white dwarf masses (0.55, 0.60, 0.65$\,M_{\odot}$) and three different initial carbon mass fractions in the envelope of the PG~1159 progenitor ($X_{\rm C}^{\rm init}=0.2,0.4,0.6$). We assume an initial envelope oxygen mass fraction $X_{\rm O}^{\rm init}=0.1$ in all cases. We set the mixing-length parameter to $\alpha=1$ and the overshoot efficiency parameter to $f_{\rm ov}=0.015$. The latter parameter was calibrated so that the 0.55$\,M_{\odot}$, $X_{\rm C}^{\rm init}=0.6$ sequence, which experiences the most significant carbon dredge-up, matches the observed DQ sequence (Figure~\ref{fig:DQseq}).\footnote{In \cite{bedard2022b}, the overshoot parameter was also calibrated based on the observed DQ sequence, but assuming instead 0.6$\,M_{\odot}$ and $X_{\rm C}^{\rm init}=0.5$, which resulted in $f_{\rm ov}=0.075$. However, \cite{bedard2022b} stressed that this value must be considered as an upper limit, because DQ stars are likely better represented by lower masses and higher $X_{\rm C}^{\rm init}$ values, which both enhance carbon dredge-up and thus reduce the amount of overshoot required to match the observed DQ sequence. This is indeed what we find in the present work: assuming 0.55$\,M_{\odot}$ and $X_{\rm C}^{\rm init}=0.6$, we obtain $f_{\rm ov}=0.015$.} This choice is motivated by the interpretation of \cite{bedard2022b} according to which DQ stars represent the most carbon-rich members of the cool helium-dominated white dwarf population, most likely as a result of their relatively low masses and high progenitor abundances (which both favour carbon pollution). Note that no hydrogen is included in our STELUM simulations. Relatively thick hydrogen envelopes can impact the dredge-up of carbon \citep[e.g., the onset of the drege-up process is delayed to $T_{\rm eff} \simeq 10{,}000\,$K if $M_{\rm H}/M_{\star}=10^{-10}$,][]{bedard2022b}, but in such cases the hydrogen content is high enough that the {\it Gaia} magnitudes are insensitive to the exact carbon photospheric abundance.

As explained in \cite{bedard2022b}, the limited range of physical conditions covered by the OPAL opacity tables \citep{iglesias1996} sets a lower limit of $T_{\rm eff} \simeq 8000\,{\rm K}$ beyond which the evolutionary calculations must be stopped to avoid extrapolating the opacity of carbon. In what follows, we linearly extrapolate the predicted $T_{\rm eff}-\log\,{\rm C/He}$ relation over a limited temperature range down to $T_{\rm eff}=7000\,$K. This allows us to study the {\it Gaia} colour--magnitude diagram down to ${\rm BP}-{\rm RP}=0.4$, which covers most of the B branch. Extrapolating over such a small range should be uncontroversial, especially given the fact that the observed $T_{\rm eff}-\log\,{\rm C/He}$ is clearly well approximated by a linear relation over that range.

\subsection{Simulation assumptions}
\label{sec:assumptions}
Simply overplotting white dwarf cooling tracks of a given mass and composition as we have done in Figure~\ref{fig:GaiaHRD_intro} is not sufficient to investigate the nature of the {\it Gaia} bifurcation as this approach fails to consider the spread in stellar mass and atmospheric composition. To take these different distributions into account, we perform Monte Carlo-based population synthesis simulations with several thousand stars. Below we describe step by step the process that we follow to generate each new star in our simulations.\\

\noindent {\bf 1. Mass:} The white dwarf mass is randomly selected from a normal distribution with a mean of $0.591\,M_{\odot}$ and a standard deviation of $0.035\,M_{\odot}$. This corresponds to the main peak of the mass distribution of DA white dwarfs warmer than 6000\,K in the \cite{kilic2020} 100~pc sample. This choice of mass distribution explicitly excludes the high- and low-mass white dwarf contributions to the overall DA mass distribution. As we have seen in Figure~\ref{fig:GaiaHRD_intro}, high-mass DAs can contaminate the B branch and the gap region. However, our goal is to explain why non-DA white dwarfs form a distinct sequence: the presence or absence of high-mass DAs below the A branch is largely irrelevant. Similarly, low-mass DAs resulting from binary evolution pathways can be ignored as they will lie above the A branch, outside the region of interest in this work. 

Since the non-DA mass distribution remains somewhat uncertain, we also use this same DA mass distribution for non-DA white dwarfs. The issue is that below $T_{\rm eff} \simeq 11{,}000\,$K the mass of non-DA white dwarfs is sensitive to the assumed hydrogen-to-helium and carbon-to-helium abundance ratios \citep{bergeron2019}. This uncertainty dissolves at higher temperatures where hydrogen traces are easier to detect and where both hydrogen and carbon traces have a negligible impact on the inferred photometric mass \citep{genest2019b} because ionized helium becomes the main electron donor (Figure~\ref{fig:electron_donor}). However, there are too few DB(A) white dwarfs in current volume-complete samples to establish a well-defined mass distribution. That said, the strong similarity between the DA and DB median masses in non-volume-complete samples \citep{bergeron2011,genest2019b,tremblay2019b} suggests that using the main peak of the DA mass distribution to model non-DA white dwarfs is likely an excellent approximation. Note also that the mass distribution of DB(A) white dwarfs falls sharply below $\simeq 0.5\,M_{\odot}$ \citep{genest2019b,tremblay2019b}, which is consistent with our choice of neglecting the low-mass contribution to the DA mass distribution as discussed above.

To obtain the {\it Gaia} magnitudes of a white dwarf with a given mass and effective temperature, we interpolate in our model atmosphere grid, using the mass--radius relation of the \cite{bedard2020} carbon--oxygen core cooling models to convert between stellar masses and surface gravities (which are used to define our model atmosphere grid).\\

\noindent {\bf 2. Effective temperature:} The effective temperature of each white dwarf is randomly selected from the temperature distribution of the 100~pc sample of \cite{kilic2020}. A Gaussian noise with mean 0 and standard deviation $300\,$K is added to smooth out the distribution and avoid generating several objects with the exact same temperature. Note that the temperature distribution is not a critical quantity in the context of this work. We are interested in reproducing the overall structure of the observed colour--magnitude diagram, which can be achieved using any reasonable temperature distribution. Indeed, it is the locations of the cooling sequences (which vary with mass and composition) that largely control the shape of the colour--magnitude diagram; the temperature distribution only determines how tightly packed stars are along these evolutionary tracks.\\

\noindent {\bf 3. Hydrogen-to-helium ratio:} 70\% of objects are assumed to keep a pure-hydrogen atmosphere throughout their evolution, having hydrogen layers thick enough that they are never mixed with the helium mantle. The exact fraction of objects that follow this channel remains uncertain, but from various studies of white dwarf spectral evolution \citep{bergeron2001,tremblay2008,limoges2015,blouin2019c,ourique2019,mccleery2020,lopez2022,caron2023} we can infer that around 70\% of white dwarfs have pure-hydrogen atmospheres at 6000\,K, a temperature by which both convective dilution and convective mixing had the opportunity to operate. Most of those objects presumably have the canonical $M_{\rm H}/M_{\star} = 10^{-4}$ hydrogen layer predicted by standard stellar evolution theory \citep{iben1984,renedo2010}.

The remaining 30\% are assumed to enter the white dwarf cooling track as hydrogen-deficient, carbon-rich PG~1159 stars as a result of a late helium-shell flash \citep{althaus2005,werner2006}.\footnote{The carbon-poor O(He) stars are also expected to produce hydrogen-deficient white dwarfs but are much rarer than their carbon-rich PG~1159 counterparts \citep{reindl2014}, so their contribution can be safely neglected for our purposes.} Among those, we suppose that two thirds have thin enough hydrogen layers to undergo convective dilution before reaching the bifurcation ($M_{\rm H}/M_{\star} \lesssim 10^{-14}$, \citealt{rolland2018}) and that one third experiences convective mixing at lower temperatures. The proportion of objects evolving through each of those two channels is justified by the fact that around 20\% of white dwarfs in the 10{,}000--15{,}000\,K range (before convective mixing can operate) have a helium-dominated atmosphere \citep{tremblay2008,cunningham2020,lopez2022}, while this number reaches roughly 30\% at lower temperatures. We stress that the exact proportion of white dwarfs following each of these evolutionary channels remains uncertain. The numbers given above should be interpreted as nothing more than our best guesses given the empirical constraints available in the literature.

For white dwarfs assumed to have undergone convective dilution, we suppose that each of them has an abundance ratio taken from a uniform distribution between $\log\,{\rm H/He}=-7$ and $-4.5$. At $T_{\rm eff} \simeq 12{,}000\,$K, still above the bifurcation but after convective dilution has taken place, DBA white dwarfs have $\log\,{\rm H/He}$ values spread between $-6.5$ and $-4.5$ \citep[Figure~14 of][see also Figure~\ref{fig:hhe_distrib}]{genest2019}. In spectroscopic surveys of DB(A) white dwarfs $\simeq 30\%$ show no detectable hydrogen \citep{koester2015,rolland2018}. To account for this minority of objects, we extend our $\log\,{\rm H/He}$ distribution down to $-7$ (the visibility limit of hydrogen at 12{,}000\,K is around $\log\,{\rm H/He}=-6.5$).  We could in principle extend this distribution still further down, but in practice this makes little difference as at such small hydrogen abundances carbon or helium takes over as the main electron donor (Figure~\ref{fig:electron_donor}) and the exact hydrogen concentration becomes inconsequential. We assume that this uniform distribution is maintained from $T_{\rm eff} = 12{,}000\,$K to $7000\,$K (the coolest temperature reached in our simulations), which is consistent with the predictions of envelope models \citep[see Figure~14 of][]{rolland2018}.

As for white dwarfs that experience convective mixing, we assume that they first have a pure-hydrogen atmosphere and then acquire a helium-dominated atmosphere, with final hydrogen-to-helium ratios taken from a uniform distribution between $\log\,{\rm H/He}=-4.5$ and $-3.0$. We assume that this transition occurs instantaneously at the mixing temperature given in Table~3 of \citet{rolland2018} and that the resulting hydrogen-to-helium ratio remains constant afterwards, which is an excellent approximation down to the $T_{\rm eff} = 7000\,$K limit of our simulations (see Figure~16 of \citealt{rolland2018} and Figure~19 of \citealt{bergeron2022}).\\

\noindent {\bf 4. Carbon abundance:} All hydrogen-deficient white dwarfs in our simulations are assumed to be polluted by a trace of carbon. The carbon-to-helium abundance ratio at their photosphere is obtained using the STELUM evolutionary calculations detailed at the end of Section~\ref{sec:stelum}. More specifically, C/He is calculated via interpolation in the $T_{\rm eff}-M_{\star}-X_{\rm C}^{\rm init}$ space spanned by the nine evolutionary tracks. For each star, $X_{\rm C}^{\rm init}$ is randomly selected from a uniform distribution between 0.2 and 0.6. The empirical $X_{\rm C}^{\rm init}$ distribution is not well constrained given the small number of known PG~1159 stars, but this uniform distribution is a good approximation of existing constraints \citep{werner2006,werner2014}. Note that with this prescription the majority of white dwarfs in our simulation do not show detectable carbon features in the optical, in qualitative agreement with observational constraints \citep{kilic2020,mccleery2020}. Indeed, Figure~\ref{fig:DQseq} shows that only stars close to the lower bound of our mass distribution and to the upper bound of the $X_{\rm C}^{\rm init}$ distribution have enough carbon at their photosphere to become DQs.

\section{Simulation results}
\label{sec:results}
\subsection{Baseline simulation: no carbon dredge-up}
\label{sec:baseline}
\begin{figure*}
    \includegraphics[width=1.7\columnwidth]{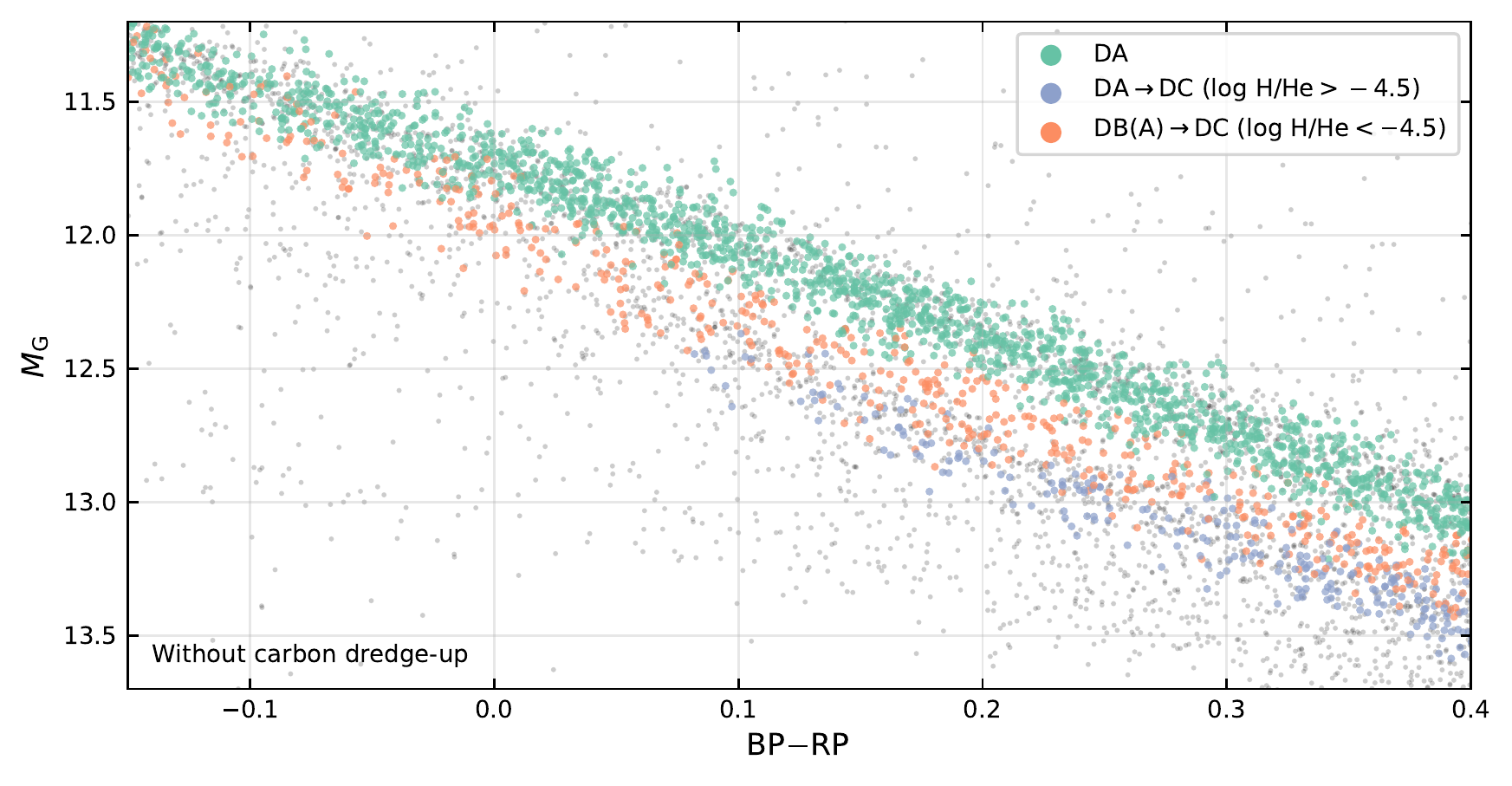}
    \caption{{\it Gaia} colour--magnitude diagram of our baseline simulation (without carbon dredge-up). The coloured circles represent individual stars in the population synthesis, and the small grey dots depict the MWDD 100~pc sample.}
    \label{fig:baseline}
\end{figure*}

In Section \ref{sec:nominal}, we will examine the results of the population synthesis simulation detailed in Section~\ref{sec:assumptions}. But before presenting this ``nominal'' simulation, we find it useful to investigate a simpler ``baseline'' simulation that neglects carbon dredge-up. Except for this omission, everything else is identical to the description given in Section~\ref{sec:assumptions}. The main output of this simulation is shown in Figure~\ref{fig:baseline}, where we compare the observed {\it Gaia} colour--magnitude diagram of the MWDD 100~pc sample (grey dots) to that produced by the simulation (coloured circles).\footnote{In what follows, we restrict our analysis to {\it Gaia} colour--magnitude diagrams as we have no reason to believe that colour--magnitude diagrams at other wavelengths (in the infrared for example) can provide useful additional information to constrain the nature of the B branch. Indeed, the main effect resulting from the addition of traces of hydrogen or traces of carbon to helium-dominated atmospheres is the same: an enhancement of He$^-$ free--free absorption. Except in the far UV, there is very little difference between a helium-dominated atmosphere with optically spectroscopically invisible traces of carbon or hydrogen.} Each simulated star is coloured according to its evolutionary channel. White dwarfs with thick hydrogen layers that maintain a pure-hydrogen atmosphere are shown in teal; hydrogen-deficient stars that have undergone convective mixing while on the B branch are shown in violet (DA$\rightarrow$DC channel); and hydrogen-deficient stars that have acquired a helium-dominated atmosphere through convective dilution before the B branch are shown in orange (DB(A)$\rightarrow$DC channel). 

If we first focus on the DA white dwarfs in Figure~\ref{fig:baseline}, we can see that the simulated A branch closely matches the observed A branch, both in terms of location and width. This suggests that our choice of mass distribution was appropriate, since the mean mass controls the vertical position of the A branch and the spread of the distribution modulates its width. Contrary to the observations, our simulation does not contain any star above the A branch. As explained in Section~\ref{sec:assumptions}, this is by design. We have not included any low-mass white dwarfs in our simulation as we are solely interested in explaining the bifurcation between the A and B branches, both populated by normal-mass white dwarfs with $\langle M_{\star} \rangle \simeq 0.6\,M_{\odot}$. Similarly, we do not attempt to replicate the Q branch, clearly visible in Figure~\ref{fig:baseline} between $M_{\rm G}=13.0$ and 13.5 and ${\rm BP}-{\rm RP}=-0.1$ and $0.3$.

\begin{figure}
    \includegraphics[width=1\columnwidth]{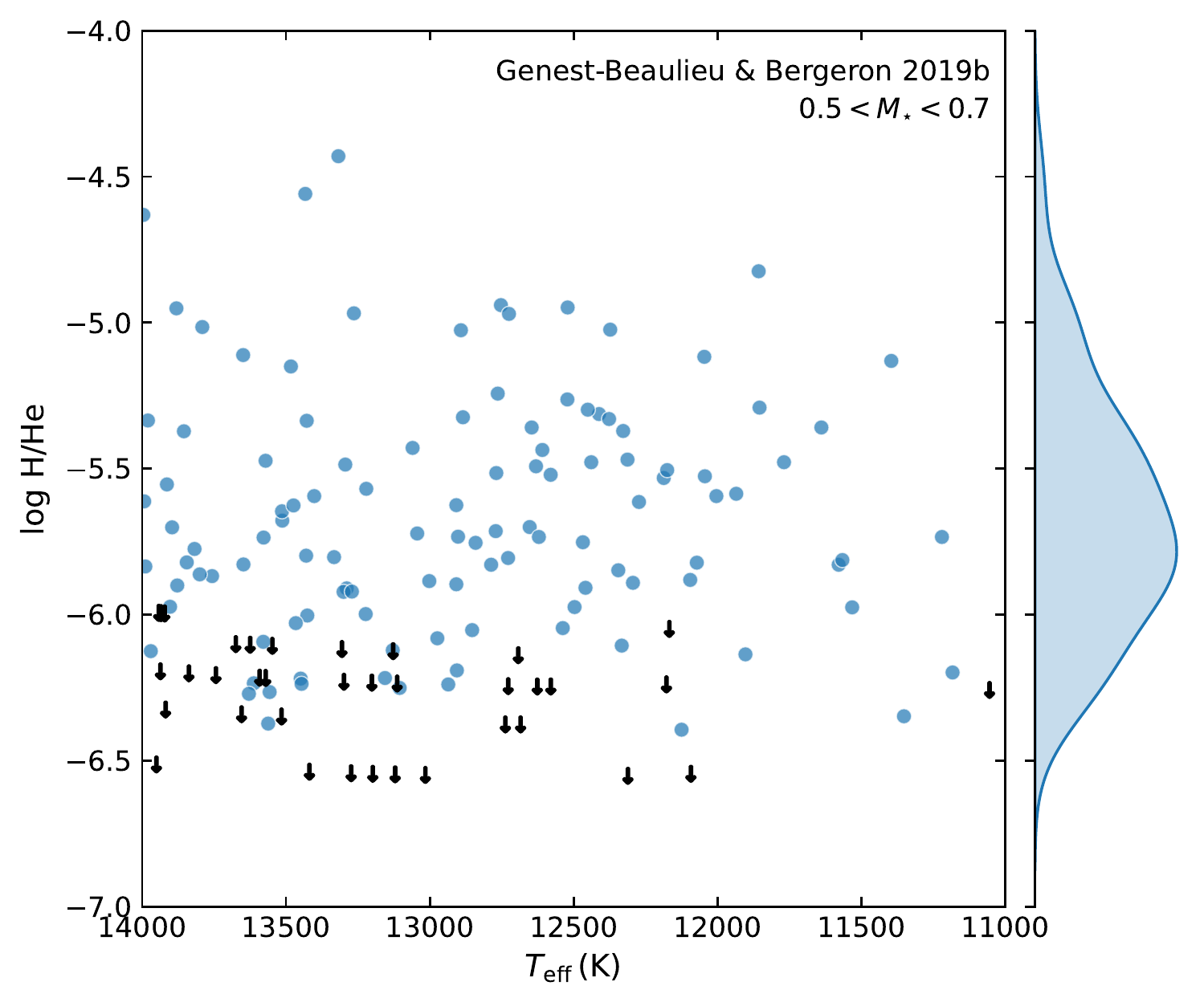}
    \caption{Hydrogen-to-helium abundance ratio in the DB(A) sample of \protect\cite{genest2019}. Circles represent DBA white dwarfs and arrows represent DB white dwarfs where only upper limits on the hydrogen abundance were determined. The distribution of $\log\,{\rm H/He}$ is shown on the right; it exclusively considers H/He measurements and ignores upper limits. Only white dwarfs with masses between 0.5 and 0.7$\,M_{\odot}$ are shown.}
    \label{fig:hhe_distrib}
\end{figure}

As expected, helium-dominated white dwarfs with large traces of hydrogen (in violet in Figure~\ref{fig:baseline}) successfully match the B branch, with none of them falling in the gap between the two branches. But the situation is completely different for objects with very small traces of hydrogen (in orange). Many of these DC white dwarfs are located right in the gap, in sharp disagreement with the observations. Naturally, the number of objects in the gap directly depends on the assumed hydrogen-to-helium ratio distribution among white dwarfs in the DB(A)$\rightarrow$DC channel. One may question whether this result simply stems from our particular choice of a uniform abundance distribution between $\log\,{\rm H/He} = -7$ and $-4.5$. Indeed, it is possible to depopulate the gap if we assume instead a non-uniform distribution with a mean closer to $-4.5$ than $-7$. However, such an assumption would go directly against the available empirical constraints on H/He. Figure~\ref{fig:hhe_distrib} shows the distribution of hydrogen-to-helium abundance ratios measured by \cite{genest2019} in their sample of DB(A) white dwarfs. In this sample, 40\% of DB(A)s with $T_{\rm eff}<14{,}000\,$K have $\log\,{\rm H/He}<-6$, a fact that is consistent with our assumption of a uniform distribution between $-7$ and $-4.5$. Any hypothetical non-uniform distribution that favours more hydrogen-rich objects would violate this constraint. Assuming that DB(A) white dwarfs maintain a constant hydrogen-to-helium abundance ratio as they cool down and become DCs (see Figure~14 of \citealt{rolland2018}), 40\% of stars that follow this channel should have $\log\,{\rm H/He}<-6$ on the B branch, an abundance that as we have seen in Figure~\ref{fig:GaiaHRD_intro} is too small to push those stars out of the gap. Therefore, the conclusion that a large fraction of the descendants of DB(A) white dwarfs end up in the gap region is robust: it is not merely an artefact of our particular choice of H/He distribution. Our baseline simulation convincingly demonstrates that the observed bifurcation cannot be explained solely by the convective dilution and mixing of residual hydrogen in helium-dominated white dwarfs.

\subsection{Nominal simulation: with carbon dredge-up}
\label{sec:nominal}
\begin{figure*}
    \includegraphics[width=1.7\columnwidth]{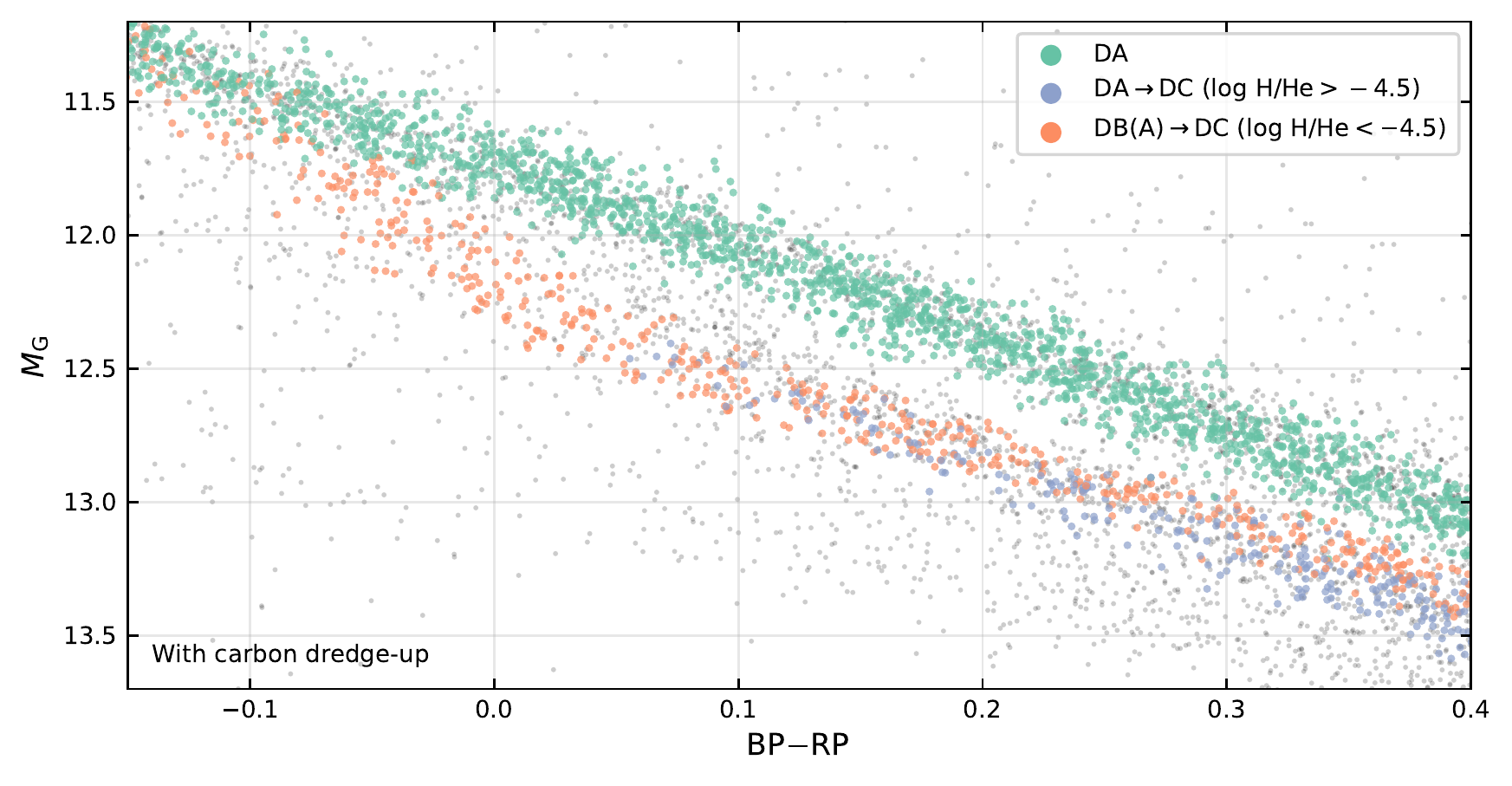}
    \caption{{\it Gaia} colour--magnitude diagram of our nominal simulation (with carbon dredge-up). The symbols have the same meaning as in Figure~\ref{fig:baseline}.}
    \label{fig:nominal}
\end{figure*}

Figure~\ref{fig:nominal} shows the results of our nominal simulation, where carbon dredge-up was included as detailed in Section~\ref{sec:assumptions}. This time, the simulation successfully produces a clear gap between the A and B branches. The addition of (mostly) optically spectroscopically invisible traces of carbon following a prescription informed by the most recent evolutionary models is sufficient to solve the problem shown in Figure~\ref{fig:baseline} and fully explain the {\it Gaia} bifurcation. The presence of carbon dredged-up from the interior compensates for the lack of hydrogen in the  descendants of DB(A) white dwarfs with $\log\,{\rm H/He} \lesssim -6$. However, solving this problem has come at a cost. Instead of separating from the A branch at ${\rm BP}-{\rm RP} \simeq 0.05$, the B branch starts at ${\rm BP}-{\rm RP} \simeq -0.05$, in clear tension with the observations. By comparing Figures~\ref{fig:baseline} and~\ref{fig:nominal}, we can infer that this behaviour is entirely due to carbon (and not to hydrogen). We will return to this problem in Section~\ref{sec:carbon}.

It is currently commonplace in the photometric analysis of helium-rich DC white dwarfs to use model atmospheres that contain a trace of hydrogen, typically $\log\,{\rm H/He} = -5$ \citep{bergeron2019,kilic2020,tremblay2020,mccleery2020,gentile2021,obrien2023,caron2023}. This is motivated by the fact that a $0.6\,M_{\odot}$ cooling sequence with $\log\,{\rm H/He} = -5$ matches the {\it Gaia} B branch (Figure~\ref{fig:GaiaHRD_intro}, \citealt{bergeron2019,kilic2020}). But if carbon pollution is the main driver of the bifurcation for a large fraction of DC white dwarfs, one may wonder whether this approximation remains justified. All things considered, we think this approach still makes sense. The impact of carbon traces on the spectral energy distribution is similar to that of hydrogen traces, because in both cases the first-order effect is an increase of the free electron density, which in turn impacts He$^-$ free-free absorption. The main effect of carbon pollution is therefore captured in carbon-free He+H atmosphere models. There is also a second-order effect due to opacity sources proper to hydrogen/carbon, but this effect is more subtle (Figure~\ref{fig:SED_demo}), especially for hydrogen/carbon abundances that remain below the visibility limit. Moreover, for such stars where invisible traces of hydrogen/carbon have a significant impact on the spectral energy distribution, there will always be an irreducible uncertainty on their atmospheric parameters associated to their unknown hydrogen/carbon abundances (except perhaps when UV spectroscopy is available). It therefore appears futile to worry about the subtle impact of carbon-specific opacities, when there is an uncertainty of $\sim 1\,$dex on the carbon abundance of a given DC star (Figure~\ref{fig:DQseq}).

\section{Discussion}
\label{sec:discussion}
\subsection{Accretion}
We have shown in Section~\ref{sec:nominal} how the predicted dredge-up of optically spectroscopically invisible amounts of carbon can naturally explain why the most hydrogen-deficient white dwarfs are located on the B branch rather than in the gap. Here, we investigate (and largely dismiss) two alternative explanations that one may invoke to explain the {\it Gaia} bifurcation.

\subsubsection{Metal accretion}
\label{sec:metals}
Metals heavier than carbon can be another source of free electrons in helium-dominated atmospheres and thus change the white dwarf's spectral energy distribution through an enhancement of the He$^-$ free--free opacity \citep{dufour2007}. These additional electron donors can in principle play the same role as carbon, pushing helium-atmosphere white dwarfs with $\log\,{\rm H/He} \lesssim -6$ out of the gap. The presence of metals heavier than carbon in warm and cool white dwarfs results from the accretion of rocky planetary material \citep{jura2014,farihi2016}. Around $15\%$ of helium-atmosphere white dwarfs show detectable absorption lines from rock-forming elements in their optical spectra \citep{mccleery2020}. This fraction is much too small to explain the {\it Gaia} bifurcation, and in any case DZs only represent a minority of objects on the B branch (Figure~\ref{fig:GaiaHRD_intro}).

Rock-forming elements should sink out of sight below the base of convection zone over timescales in the millions of years for helium-atmosphere white dwarfs on the B branch \citep{koester2020,heinonen2020}. Therefore, stars that currently show metal absorption lines in their spectra are thought to be actively accreting or to have recently (over the past millions of years) accreted rocky material. This implies that some fraction of the remaining $\sim 85\%$ of helium-atmosphere white dwarfs must harbour some spectroscopically invisible amount of metals acquired during a past accretion event. So while DZs alone cannot account for the {\it Gaia} bifurcation, maybe DC stars with non-zero but spectroscopically invisible metal abundances can?

\begin{figure}
    \includegraphics[width=\columnwidth]{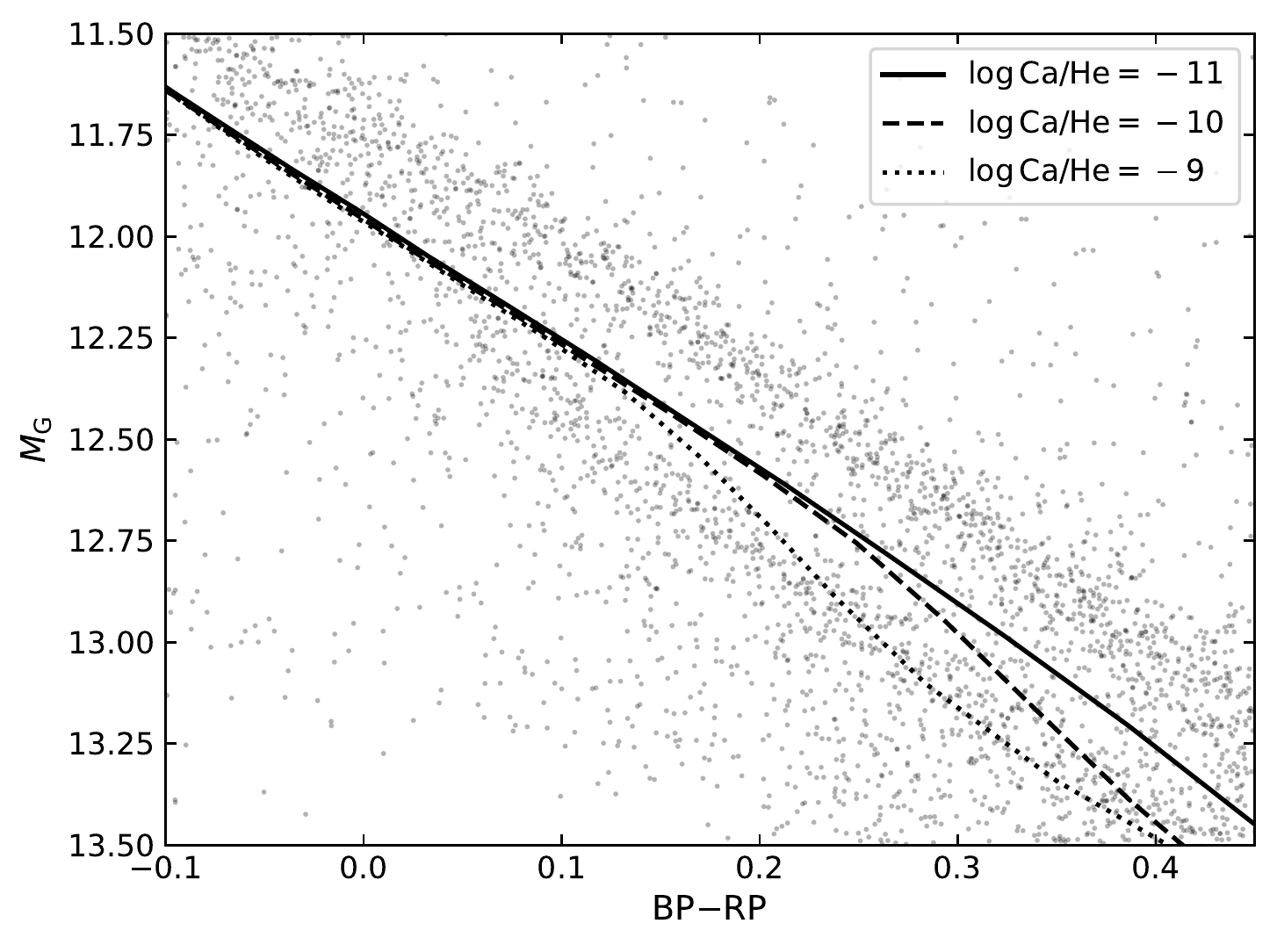}
    \caption{Cooling sequences of $0.6\,M_{\odot}$ helium-atmosphere white dwarfs with different pollution levels. The legend gives the calcium number abundance ratio; all other elements up to copper are included assuming chondritic abundance ratios with respect to calcium \citep{lodders2003}. The grey dots represent the MWDD 100~pc sample.}
    \label{fig:DZ}
\end{figure}

To answer this question, we show in Figure~\ref{fig:DZ} cooling sequences of $0.6\,M_{\odot}$ helium-atmosphere white dwarfs with different amounts of polluting material. These model atmospheres were presented in \cite{coutu2019}; all metals from calcium to copper are included assuming chondritic relative abundance ratios. Clearly, $\log\,{\rm Ca/He}>-10$ is needed for the cooling track to coincide with the B branch, and even then it completely misses it at ${\rm BP}-{\rm RP} \lesssim 0.2$. This result is sufficient to invalidate the hypothesis that metal accretion is a viable explanation for the {\it Gaia} bifurcation. At $\log\,{\rm Ca/He}>-10$, the calcium H \& K lines are very strong and these stars would have been classified as DZs \citep{dufour2007,hollands2017,coutu2019,blouin2020b}, in contradiction with the observation that the majority of stars on the B branch are DCs or DQs (Figure~\ref{fig:GaiaHRD_intro}). We can therefore safely reject the idea that metal accretion is an important cause of the {\it Gaia} bifurcation.

\subsubsection{Hydrogen accretion}
\label{sec:Haccretion}
So far, we have assumed that the hydrogen-to-helium ratio of a convectively diluted atmosphere remains constant throughout the B branch. This is consistent with our understanding of the internal processes that shape spectral evolution \citep{rolland2018,bedard2023}, but it is possible that an external process increases the white dwarf's hydrogen content. In particular, accretion from the interstellar medium \citep{macdonald1991,voss2007} or from water-bearing planetesimals \citep{farihi2013,raddi2015,gentile2017,izquierdo2021} can provide additional hydrogen to the star as it cools down. One can imagine a scenario where this external supply of hydrogen is the source of free electrons that pushes the helium-atmosphere white dwarfs out of the gap and toward the B branch in the {\it Gaia} colour--magnitude diagram. 

For this mechanism to be the main cause of the bifurcation, it needs to conform to the observational constraint that the B branch is constituted in large part of DC white dwarfs without any detectable Balmer lines. (This is not to say that all stars on the B branch do not show Balmer lines: there are some DZA and helium-rich DA white dwarfs.) To test the plausibility of this scenario, we have performed a simulation identical to our baseline simulation (no carbon dredge-up, Figure~\ref{fig:baseline}), but where the hydrogen-to-helium ratio of every helium-atmosphere white dwarf cannot be smaller than the hydrogen visibility limit given in Figure~4 of \cite{coutu2019}. If the hydrogen abundance selected from our random distribution produces a helium-rich DA white dwarf, we leave its H/He ratio unchanged; if it produces a DC, we raise its H/He right to the hydrogen detection threshold. This represents an extreme (and, admittedly, highly unlikely) scenario, in the sense that all DC white dwarfs are just on the verge of displaying detectable Balmer lines. Adding any more hydrogen than that would transform those stars into helium-rich DAs, which is not a viable option as from observations we know that DC stars vastly outnumber helium-rich DAs \citep{caron2023}.

\begin{figure*}
    \includegraphics[width=1.7\columnwidth]{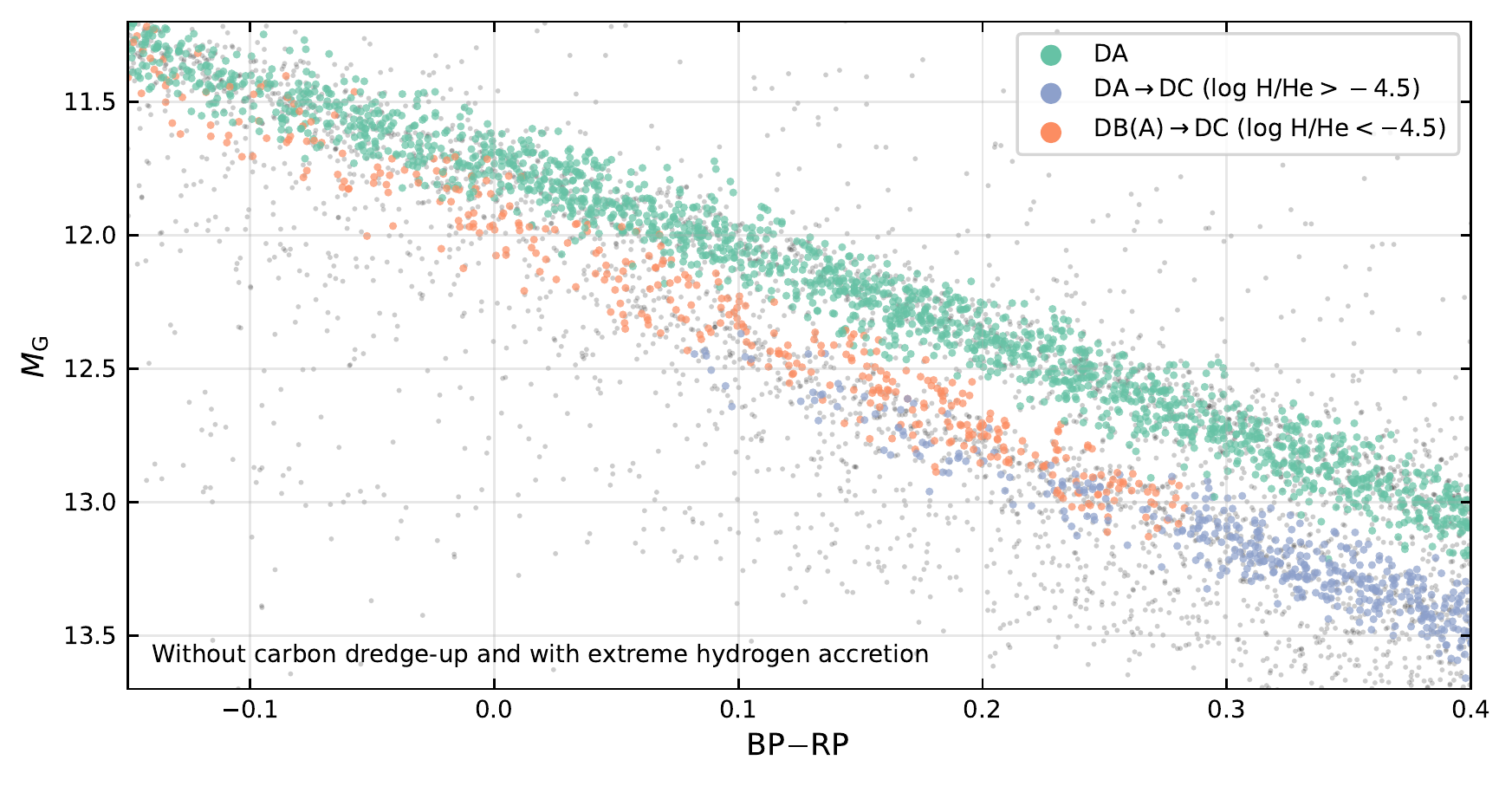}
    \caption{{\it Gaia} colour--magnitude diagram of our baseline simulation (without carbon dredge-up) and with the extreme hydrogen accretion scenario described in Section~\ref{sec:Haccretion}. The symbols have the same meaning as in Figure~\ref{fig:baseline}.}
    \label{fig:accretion}
\end{figure*}

The results of this simulation are shown in Figure~\ref{fig:accretion}. It can be seen that even with this far-fetched scenario there are still many objects in the gap. This is especially true at ${\rm BP}-{\rm RP} < 0.2$, where hydrogen is more easily detectable and a smaller quantity can be hidden in a featureless DC star (only $\log\,{\rm H/He} \lesssim -5.5$). In all likelihood, hydrogen accretion is weaker than the extreme scenario presented here, and we can therefore confidently reject the idea that it could explain the paucity of objects in the gap. This solution is simply incompatible with the large number of DC white dwarfs on the B branch: more hydrogen would be required (in this temperature and composition regime, stars are pushed out of the gap and toward the B branch in a monotonic fashion as more hydrogen is added).

\subsection{On the origin of the B branch prolongation at \boldmath${\rm \bf BP}-{\rm \bf RP} < 0.1$\unboldmath}
\label{sec:carbon}
Having ruled out accretion as a viable explanation for the {\it Gaia} bifurcation, we now return to our nominal simulation with carbon dredge-up. As we have seen, the dredge-up of invisible traces of carbon is thought to be inevitable for most helium-atmosphere white dwarfs and can naturally explain the split in the colour--magnitude diagram (Figure~\ref{fig:nominal}). However, current models also predict a prolongation of the bifurcation at ${\rm BP} - {\rm RP} < 0.1$ that is at odds with the observational data. Here, we investigate two potential solutions to this problem.

\subsubsection{Inappropriate model atmospheres}
A first possible explanation is that current model atmospheres incorrectly predict the spectral energy distributions of carbon-polluted white dwarfs in this temperature range ($9500\,{\rm K} \lesssim T_{\rm eff} \lesssim 12{,}000\,$K). If this is the case, it seems unlikely that the ionization balance of carbon, which controls the free electron density and the strength of He$^-$ free--free absorption, is to blame. While assessing the ionization equilibrium is certainly a hard problem in cool ($T_{\rm eff} \lesssim 6000\,$K) helium-dominated atmospheres \citep{kowalski2007,blouin2018}, there is no known uncertainty for those warmer objects where the photospheric densities remain in the regime of applicability of the ideal Saha ionization equation \citep[see Figure~16 of][]{saumon2022}. 

We have seen in Figure~\ref{fig:SED_demo} that while He$^-$ free--free enhancement is the main reason for the change in the spectral energy distribution of lightly carbon-polluted atmospheres, the carbon opacities themselves are not totally negligible (compare the dotted and dashed lines). Strong transitions in the UV lead to flux redistribution at longer wavelengths that affect the {\it Gaia} magnitudes of the star. DQ white dwarfs often show both atomic and molecular carbon features in the $9500\,{\rm K} \lesssim T_{\rm eff} \lesssim 11{,}000\,$K temperature range. Interestingly, current atmosphere models fail to simultaneously match both types of absorption features \citep[Figure~14 of][]{coutu2019}. These shortcomings have been blamed on uncertainties in the carbon UV opacities (possibly due to unreliable atomic data or inappropriate line-broadening prescriptions), which could lead to an offset of the temperature, mass and abundance scales of DQ white dwarfs \citep{coutu2019}. It is therefore tempting to attribute the bifurcation prolongation in the $9500\,{\rm K} \lesssim T_{\rm eff} \lesssim 12{,}000\,$K range to this known problem. We think this is a promising avenue, but there is at least one complication with this hypothesis. As argued by \cite{coutu2019}, if the UV opacities are off, they most likely affect all carbon-polluted atmospheres, not just those in the $9500\,{\rm K} \lesssim T_{\rm eff} \lesssim 11{,}000\,$K range. We happen to notice this problem in this specific temperature regime because it is only then that molecular and atomic carbon features simultaneously appear in the optical, but there is a priori no reason to believe that the impact of any opacity-related problem would be restricted to this narrow temperature range. Indeed, this discrepancy is also noticeable at lower effective temperatures for the few stars that have the UV data required to simultaneously detect atomic and molecular carbon below 9500\,K \citep[Figure 3.3 of][]{dufour2011}. In other words, it is not obvious that a different opacity model could remediate the B branch prolongation without creating new problems at lower and/or higher temperatures.

\subsubsection{A more sudden dredge-up}
 \begin{figure*}
    \includegraphics[width=1.7\columnwidth]{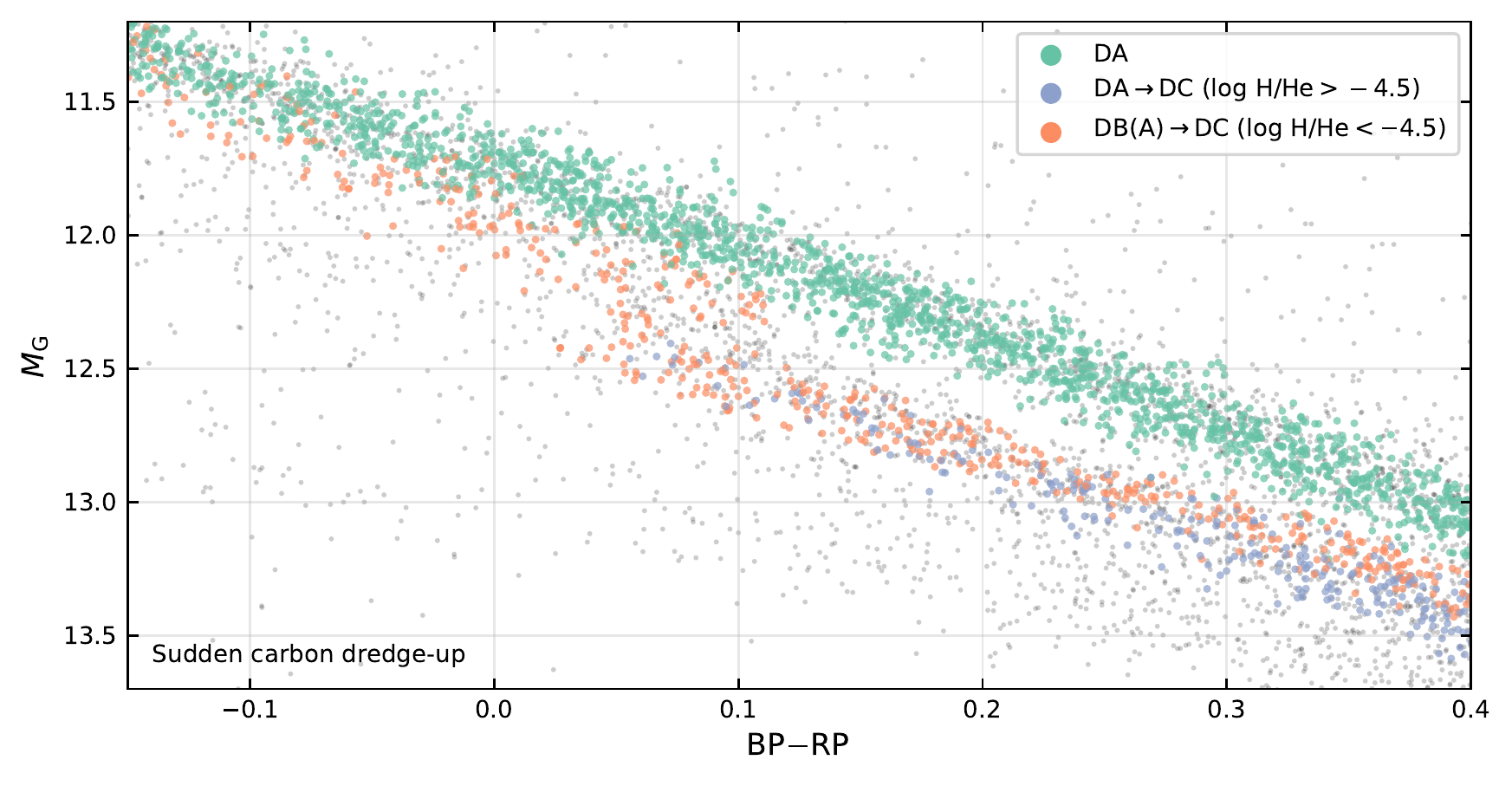}
    \caption{{\it Gaia} colour--magnitude diagram of our simulation with a sudden dredge-up of carbon at $T_{\rm eff}=10{,}000\,$K (see text for details). The symbols have the same meaning as in Figure~\ref{fig:baseline}.}
    \label{fig:sudden_du}
\end{figure*}

Another possible explanation is that the problem comes from upstream and is caused instead by the evolutionary models. In particular, the tension would be eased if helium-atmosphere white dwarfs in the $9500\,{\rm K} \lesssim T_{\rm eff} \lesssim 12{,}000\,{\rm K}$ range had less carbon in their atmospheres. To demonstrate this point, we show in Figure~\ref{fig:sudden_du} the results of a simulation identical to our nominal simulation (Figure~\ref{fig:nominal}), but where the carbon abundance is arbitrarily forced to zero above $T_{\rm eff}=10{,}000\,$K. Evidently, this scenario matches the observations remarkably well. The experiment shown in Figure~\ref{fig:sudden_du} is a priori artificial and without a strong physical underpinning. \cite{bedard2022b} presented evolutionary sequences that exhibit a sudden dredge-up at $T_{\rm eff} \simeq 10{,}000\,$K (see their Figure~14), but this scenario requires a large amount of hydrogen ($M_{\rm H}/M_{\star} \simeq 10^{-10}$). It is therefore only relevant to the DA$\rightarrow$DC channel and not to the DB(A)$\rightarrow$DC channel, which is the one responsible for the prolongation of the B branch at ${\rm BP}-{\rm RP}<0.1$.

 That said, the scenario presented in Figure~\ref{fig:sudden_du} is entirely compatible with existing empirical constraints. In this temperature range, small traces of carbon are difficult to spectroscopically detect in the optical, and there are very few confirmed DQs above $T_{\rm eff} = 10{,}000\,$K (Figure~\ref{fig:DQseq}).\footnote{The rare DQ stars that do show optical carbon features in this temperature range have high masses and are not thought to have evolved through the canonical PG~1159--DO--DB--DQ channel \citep{coutu2019,kawka2023}.} UV data can in principle be used to measure carbon levels that are undetectable in the optical, but in practice there is currently too little data to constrain the evolution of the carbon abundance in any meaningful way at $T_{\rm eff} > 10{,}000\,$K \citep{bedard2022b}. All things considered, both the nominal and sudden dredge-up scenarios are compatible with DQ observations.

A sudden dredge-up is certainly a possibility from an empirical perspective, but is there any physical argument in favour of this scenario? A first place to look is the assumed overshoot parameter at the base of the helium convection zone. A weaker overshooting would delay the dredge-up process, but the overshoot parameter is already very small in our simulations ($f_{\rm ov}=0.015$, see Section~\ref{sec:stelum}) so there is too little room on this front to significantly delay the onset of carbon dredge-up. The other quantity that controls the start of the dredge-up process is the slope of the carbon diffusion tail in the deep interior. A steeper diffusion tail would demand a deeper helium convection zone (and hence a cooler $T_{\rm eff}$) before carbon can be transported to the photosphere. As it turns out, the slope of this diffusion tail remains uncertain.

\begin{figure}
\includegraphics[width=\columnwidth]{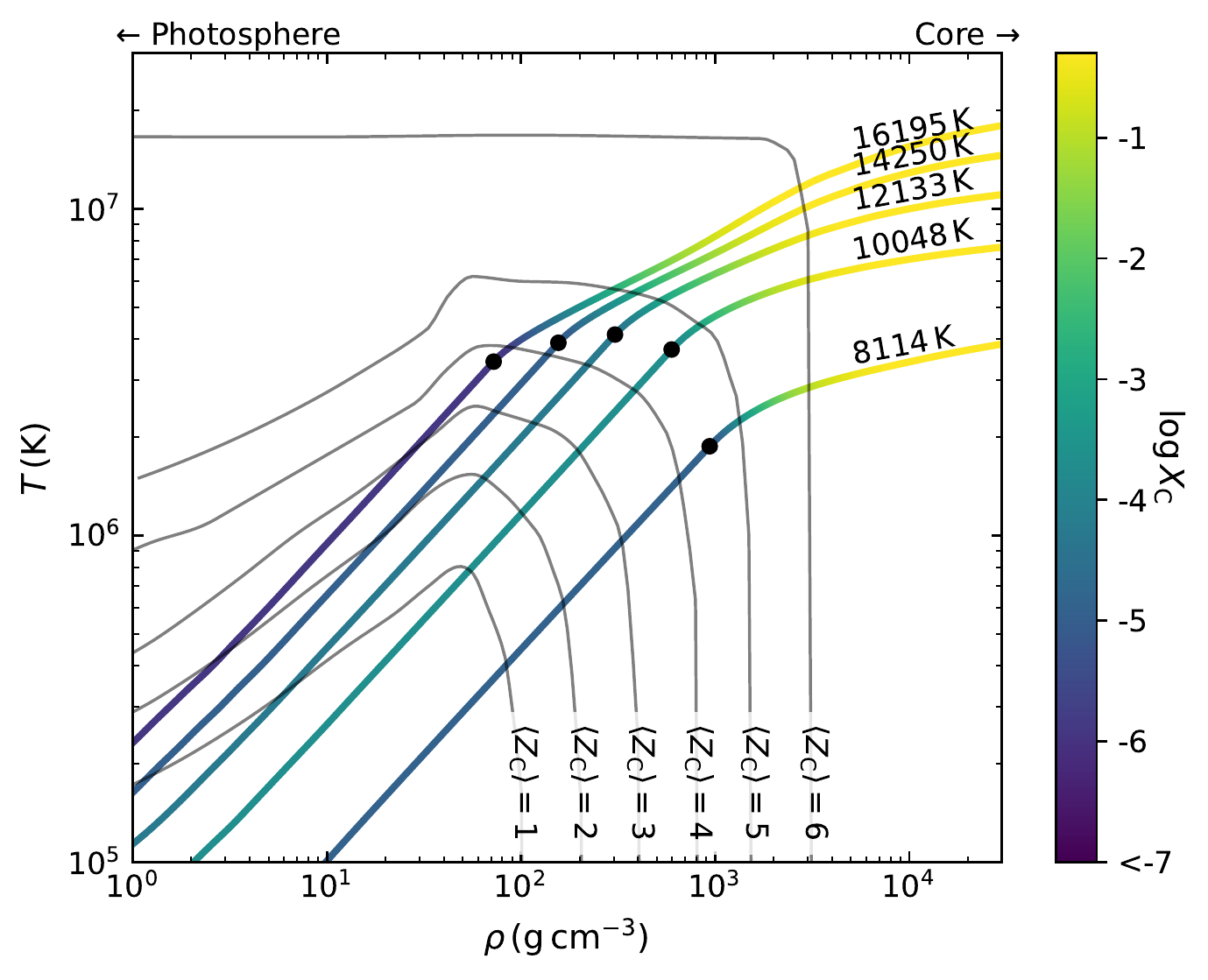}
    \caption{Density--temperature profiles of a $0.55\,M_{\odot}$ hydrogen-free white dwarf with $X_{\rm C}^{\rm init}=0.6$. Five different stages in the evolution of the star are shown; in each case the effective temperature is given next to the corresponding profile. The colour scale indicates the carbon mass fraction, with $50\%$ carbon in the core and a much smaller trace in the envelope. Black circles indicate the bottom of the helium convection zone, where carbon is dredged-up from the diffusion tail. Also shown are isocontours of the ionization state of carbon as calculated in STELUM (thin grey lines).}
    \label{fig:rhoT}
\end{figure}

As described by \cite{pelletier1986}, the slope of the carbon diffusion tail is sensitive to the ionization state of carbon. This is unfortunate because the physical conditions in this region of the star (Figure~\ref{fig:rhoT}) are notoriously challenging for equation of state models. In STELUM, the equation of state of the He+C mixture in the envelope is modelled using an additive volume rule. The carbon equation of state is taken from  \cite{fontaine1977,fontaine2001}, a simple heuristic ionization model that cannot be considered reliable in this warm dense matter regime where pressure ionization plays a key role. In fact, for conditions corresponding to $T = 2 \times 10^6\,{\rm K}$ and $\rho = 100\,{\rm g}\,{\rm cm}^{-3}$, the \cite{fontaine1977,fontaine2001} carbon EOS predicts a mean ionization state of $\langle Z_{\rm C} \rangle \simeq 3$ (Figure~\ref{fig:rhoT}), while the more sophisticated pseudo-atom molecular dynamics \citep[PAMD,][]{starrett2013,starrett2014,saumon2020} model yields $\langle Z_{\rm C} \rangle \simeq 4.5$ (D.~Saumon, private communication). But this large discrepancy does not even tell the whole story. In the star, carbon is only a trace species in an otherwise pure-helium plasma: what we really need to know is the ionization state of carbon in a bath of helium, not in a pure carbon plasma as given by the \cite{fontaine1977,fontaine2001} model.\footnote{These limitations are not specific to the equation of state implemented in STELUM. They are also inherent to the equation-of-state models of MESA \citep{jermyn2023}, LPCODE \citep{camisassa2017} and BASTI \citep{salaris2022}.}

Assessing the ionization state of mid-$Z$ elements in warm dense helium is a well-known problem in white dwarf physics \citep{saumon2022}. It lies at the heart of the calculation of the diffusion timescales that are needed to interpret the photospheric abundances of polluted white dwarfs in the context of planetesimal accretion \citep{bauer2019}. Current calculations still rely on simple continuum-lowering \citep{paquette1986b,dupuis1992,fontaine2015} or Thomas--Fermi \citep{stanton2016,koester2020} models of pressure ionization. \cite{heinonen2020} have shown that a more sophisticated average-atom model is in strong disagreement with these simple prescriptions (see in particular their Figures~2 and~3). Unfortunately, such advanced ionization models are not publicly available at the moment, and therefore no good option for the treatment of the partial ionization of carbon in dense helium currently exists.

\begin{figure}
     \includegraphics[width=\columnwidth]{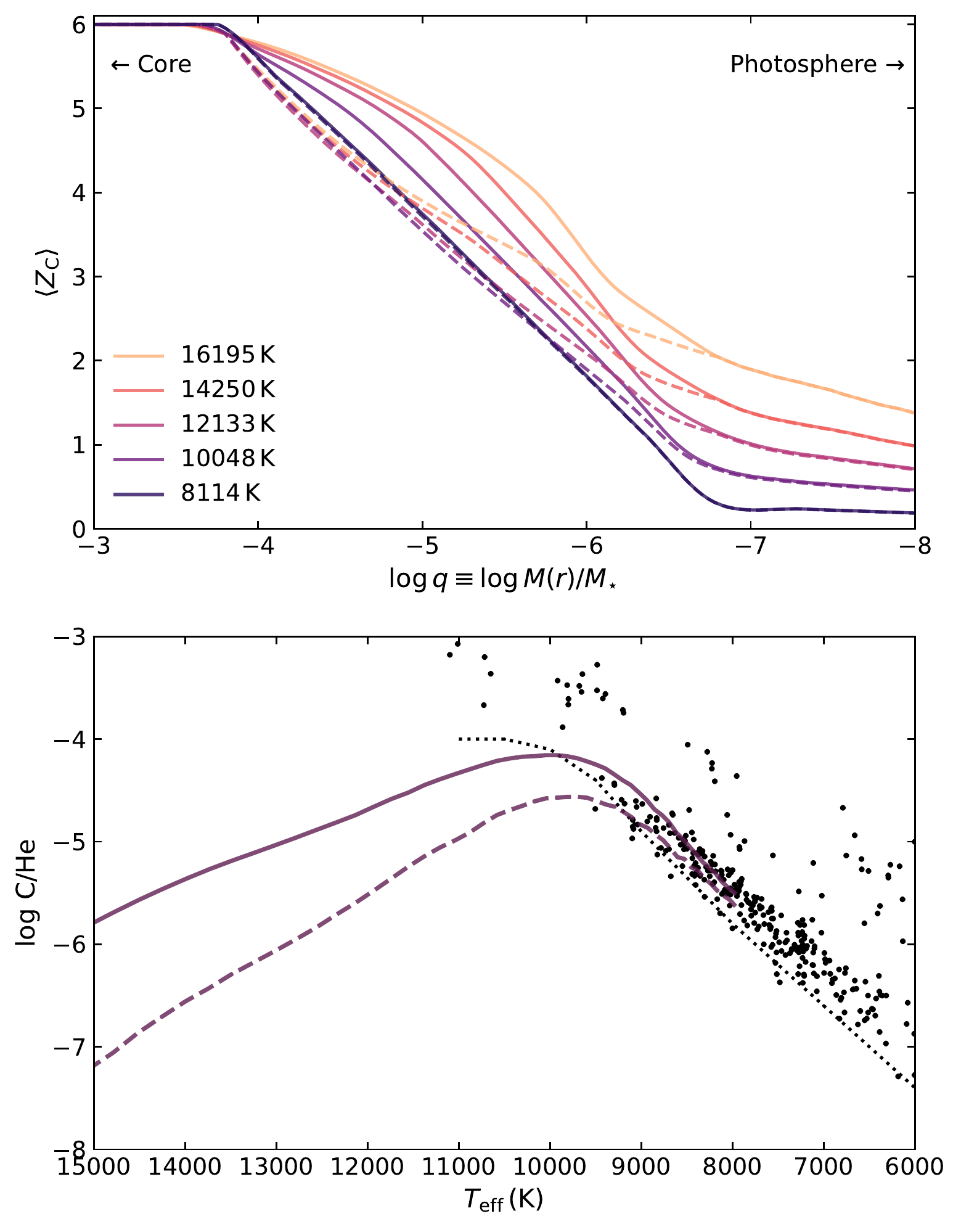}
    \caption{Top: Ionization state of carbon as a function of the mass coordinate for the same models as in Figure~\ref{fig:rhoT} (solid lines; $0.55\,M_{\odot}$, $X_{\rm C}^{\rm init}=0.6$, no hydrogen) and for models with an arbitrarily altered carbon ionization model (dashed lines). Bottom: Carbon abundance as a function of effective temperature for our nominal model (solid line) and for a model with the altered carbon ionization model shown in the top panel (dashed line). The carbon visibility limit (dotted line) and measured DQ parameters (black circles) are also shown for reference.}
    \label{fig:ZC}
\end{figure}

In the absence of an improved equation of state, we have performed a numerical experiment to investigate the sensitivity of carbon dredge-up on the carbon ionization state in the deep interior. For this simple test, we pick the evolutionary sequence that maximizes the quantity of dredged-up carbon (the $0.55\,M_{\odot}$ and $X_{\rm C}^{\rm init}=0.6$ case). We have arbitrarily changed the ionization state of carbon in the envelope to create a smoother transition from $\langle Z_{\rm C} \rangle=1$ to $\langle Z_{\rm C} \rangle=6$ as we move toward the center of the star (top panel of Figure~\ref{fig:ZC}). We stress that there is no physical basis for this particular prescription: the goal of this exercise is merely to explicitly demonstrate the sensitivity of the amount of dredged-up carbon on the equation of state. In the bottom panel of Figure~\ref{fig:ZC}, we show how the sequence with the altered ionization model (dashed lines) compares to the fiducial sequence (solid line) in terms of the $T_{\rm eff} - {\rm C/He}$ relation. The adjustments to the ionization state of carbon result in much less carbon at high temperature before the DQ sequence, precisely what would be required to alleviate the discrepancy identified in Figure~\ref{fig:nominal}. It is our hope that this will help motivate future work on the ionization state of carbon in warm dense helium.

\section{Conclusion}
\label{sec:conclusion}
We have performed population synthesis simulations to elucidate the exact nature of the bifurcation in the white dwarf cooling sequence of the {\it Gaia} colour--magnitude diagram. We showed that traces of hydrogen in helium-dominated atmospheres cannot fully account for the observed separation between hydrogen- and helium-atmosphere white dwarfs. Given our current understanding of white dwarf chemical evolution and empirical constraints from DB(A) white dwarfs, a large fraction of helium-atmosphere white dwarfs have too little hydrogen to recreate the observed bifurcation. We argued that the natural solution to this problem is the dredge-up of optically undetectable amounts of carbon from the deep interior. DQ white dwarfs are too rare to explain the {\it Gaia} bifurcation, but a large population of featureless though carbon-polluted ``DQ-manqué'' white dwarfs is predicted to exist. Based on the most up-to-date model atmospheres and evolutionary calculations, we demonstrated that the expected carbon abundances in those stars is sufficient to fully account for the observed bifurcation. The only caveat with this scenario is the prediction of an over-extended bifurcation at high temperatures, which is plausibly due to limitations of currently available equation-of-state models. Moreover, we have investigated alternative scenarios involving the accretion of metals or hydrogen, but rejected them on the basis that, contrary to carbon, the quantity of polluting material required to reproduce the bifurcation would necessarily lead to detectable spectroscopic features.

We highlight two promising avenues for future work. The first is observational. While the small carbon traces of DQ-manqués are invisible in the optical, they can often be detected in the UV. Programs focusing on acquiring UV spectra of DC and cool DB white dwarfs would likely uncover this hidden population, thereby providing precious anchor points for evolutionary models and testing our claim that traces of carbon invisible in the optical are key to explaining the {\it Gaia} bifurcation. According to our scenario, the majority of $0.6\,M_{\odot}$ non-DA white dwarfs at $T_{\rm eff} \simeq 10{,}000\,$K should have photospheric carbon abundances in excess of $\log\,{\rm C/He}=-7$. This is a level of pollution that has already been shown to be detectable in the UV for stars in this $T_{\rm eff}$ regime \citep[e.g.,][]{weidemann1995,xu2013}. Therefore, a small UV spectroscopic survey of $\sim 10$ DC white dwarfs in this temperature range would be sufficient to decisively test the main claim of this work. The second avenue for future research concerns the equation of state of carbon in warm dense helium. As we have shown, this is a crucial quantity in the modelling of the carbon dredge-up process, not only for DQ-manqués but also for regular DQ white dwarfs.

\section*{acknowledgements}
We are grateful to Didier Saumon for insightful discussions on the equation of state of partially ionized carbon and for sharing unpublished PAMD results. We also thank Patrick Dufour for useful discussions on the UV--optical discrepancies of carbon abundances in DQ white dwarfs. We thank the referees for useful comments that have improved this work. SB is a Banting Postdoctoral Fellow and a CITA National Fellow, supported by the Natural Sciences and Engineering Research Council of Canada (NSERC). This work has made use of the Montreal White Dwarf Database \citep{dufour2017}. This work has benefited from discussions at the KITP program ``White Dwarfs as Probes of the Evolution of Planets, Stars, the Milky Way and the Expanding Universe'' and was supported in part by the National Science Foundation under Grant No. NSF PHY-1748958. AB and PET
have received funding from the European Research Council (ERC)
under the European Union’s Horizon 2020 research and innovation programme (Grant agreement No. 101020057).

\section*{Data availability}
All observational data used in this work is publicly available on the {\it Gaia} archive and Montreal White Dwarf Database.

\bibliographystyle{mnras}
\bibliography{references}

\bsp
\label{lastpage}

\end{document}